\documentclass[12pt]{article}
\usepackage{times}
\usepackage{geometry}
\usepackage{graphicx}
\usepackage{caption}
\usepackage{subcaption}
\usepackage{titlesec}

\titleformat*{\section}{\large\bfseries}
\titleformat*{\subsection}{\normalsize\bfseries}
\titleformat*{\subsubsection}{\large\bfseries}
\titleformat*{\paragraph}{\large\bfseries}
\titleformat*{\subparagraph}{\large\bfseries}

\usepackage[utf8]{inputenc}
\usepackage{enumitem,amssymb}
\usepackage{ragged2e}
\newlist{thematic}{itemize}{8}
\setlist[thematic]{label=$\square$}
\usepackage{pifont}

\usepackage{color}
\definecolor{black}{rgb}{0,0,0}
\definecolor{red}{rgb}{1.0,0,0}

\begin{document}
\thispagestyle{empty}
\raggedright

{\bf Astro2020 APC White Paper: Activities and Projects}\\ 
\vspace{5px}
{\bf Commensal, Multi-user Observations with an Ethernet-based Jansky Very Large Array}\\
\vspace{5px}
\normalsize
\textbf{Thematic Areas:} Activities, Projects, Instrumentation, Ground-based Facilities, Cosmology and Fundamental Physics, Planetary Systems\\
\vspace{5px}

\textbf{Principal Author:}
Name: Dr. Jack Hickish
 \linebreak						
Institution: SETI Institute / University of California, Berkeley
 \linebreak
Email: jackh@berkeley.edu\\
\vspace{5px}
\textbf{Co-authors:} {Tony Beasley (National Radio Astronomy Observatory), Geoff Bower (Academia Sinica Institute of Astronomy and Astrophysics), Sarah Burke-Spolaor (West Virginia University), Steve Croft (University of California, Berkeley), Dave DeBoer (SETI Institute / University of California, Berkeley), Paul Demorest (National Radio Astronomy Observatory), Bill Diamond (SETI Institute), Vishal Gajjar (University of California, Berkeley), Casey Law (California Institute of Technology), Joseph Lazio (Jet Propulsion Laboratory, California Institute of Technology), Jason Manley (South African Radio Astronomy Observatory), Zsolt Paragi (Joint Institute for VLBI ERIC (JIVE)), Scott Ransom (National Radio Astronomy Observatory), Andrew Siemion (SETI Institute / University of California, Berkeley / Radboud University / University of Malta) }\\
\vspace{5px}  
\textbf{Abstract:} Over the last decade, the continuing decline in the cost of digital computing technology has brought about a dramatic transformation in how digital instrumentation for radio astronomy is developed and operated. In most cases, it is now possible to interface consumer computing hardware, e.g. inexpensive graphics processing units and storage devices, directly to the raw data streams produced by radio telescopes. Such systems bring with them myriad benefits: straightforward upgrade paths, cost savings through leveraging an economy of scale, and a lowered barrier to entry for scientists and engineers seeking to add capabilities or features to a new or existing instrument. Additionally, certain properties of these systems allow fundamentally new ways of looking at the standard operational model of a radio observatory. The typical data-interconnect technology used with general-purpose computing hardware -- Ethernet -- naturally permits multiple subscribers to a single raw data stream. Taking advantage of this feature, multiple science programs can be conducted in parallel, processing the data stream in very different ways. When combined with broad bandwidths and wide primary fields of view, radio telescopes become capable of achieving many science goals simultaneously. Moreover, because many science programs are not strongly dependent on observing cadence and direction (e.g. searches for extraterrestrial intelligence and radio transient surveys), these so-called ``commensal'' observing programs can dramatically increase the scientific productivity and discovery potential of an observatory. In this whitepaper, we detail a project to add an Ethernet-based commensal observing mode to the Jansky Very Large Array (VLA), and discuss how this mode could be leveraged to conduct a powerful program to constrain the distribution of advanced life in the universe through a search for radio emission indicative of technology.  We also discuss other potential science use-cases for the system, and how the system could be used for technology development towards next-generation processing systems for the Next Generation VLA (ngVLA).

\pagebreak
\pagenumbering{arabic} 

\section{Key Science Goals and Objectives}
The certainty with which we now know that our galaxy, and indeed the universe, is awash in oases for life has raised the question of whether there is life beyond Earth to one of the most profound scientific prospects of this century \cite{NAP25252}. Among the methods for probing this topic is the search for evidence of advanced, technologically capable life elsewhere in the universe $-$ the search for extraterrestrial intelligence (SETI) $-$ or more recently the search for ``technosignatures''.  Relative to other methods of searching, e.g. the search for spectroscopic signatures of terrestrial-like biology in exoplanet atmospheres \cite{mercedes2020}, searches for evidence of terrestrial-like technology probe vastly larger volumes of space, and vastly more targets \cite{wright2020techno, jlm2020}. The relative probability of a successful detection for these complementary methods thus comes down to the propensity with which life evolves a technological capacity, and importantly, the length of time that technology continues to be detectable \cite{2018PNAS..115E9755G}.

\begin{figure}
    \centering
    \includegraphics[width=\textwidth]{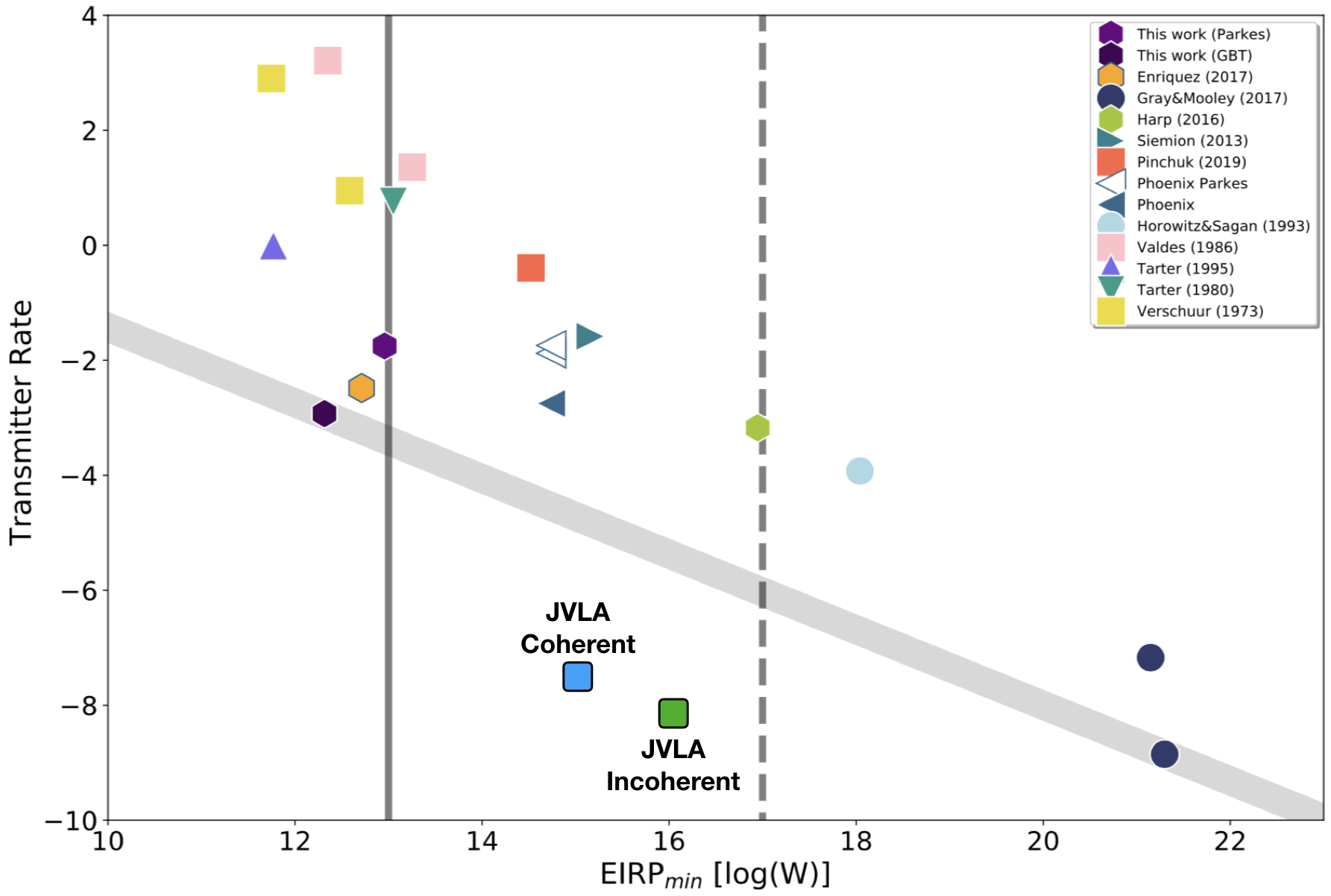}
    \caption{Adapted from \cite{Price:2019tg}, the transmitter rate limit, $(N_{\rm{star}}(\frac{\nu_{total}}{\nu_{\rm{center}}}))^{-1}$, is plotted on logarithmic axes against the minimum detectable power, EIRP$_{\rm{min}}$, based on the distance to the farthest star in the sample, for two VLA VLASS commensal SETI observing modes alongside limits achieved by other recent searches. Points toward the bottom of this plot represent surveys with large numbers of stellar targets and large fractional bandwidth; points toward the left represent surveys where sensitivity is higher and distance to targets is lower. The solid and dashed vertical lines represent the EIRP of the Arecibo planetary radar, and total Solar insolation, respectively.  A transmitter rate of 1 would be an occurrence rate of 1 narrow band sinusoid per star, per GHz, at a center frequency of 1 GHz.  For the incoherent case, we assume a survey of 250M stars out to a maximum distance of 825pc in a single epoch using an incoherent summation of the individual station data.  For the coherent case, we assume a phased-array system of 50 beams observing 50 FGK stars per primary beam with a maximum distance of 825pc.  In both cases, we assume a single epoch of VLASS with observing strategy parameters as described in \cite{Lacy:2019uf}, including a typical usable RFI-free band of 1500 MHz.  For the SETI search, we assume 1 Hz channelization and a 10$\sigma$ detection threshold.}
    \label{fig:seti_rates}
\end{figure}

Radio emission at cm-wavelengths is a particularly attractive tracer of technology, due to its relative ease of generation and reception (based on our human experience with technology), the relative transparency of the interstellar and intergalactic media to these wavelengths and the degree to which the spectral and temporal properties of artificial radio emission distinguish it from natural sources \cite{jlm2020}. A commensal radio SETI system on the Jansky Very Large Array (VLA), as described in Section \ref{sec:tech}, would permit a wide variety of searches for radio technosignatures, limited only by pointing and tuning configuration chosen by the primary telescope user. As an illustrative example of how this system could be used, in Figure \ref{fig:seti_rates} we compare the limits on narrow-band ($<$ 1 Hz) radio emitters attainable in a 32 month (single epoch) campaign commensal with the VLA Sky Survey (VLASS) \cite{Lacy:2019uf}.  As shown by the Figure and described in the caption, a commensal program operating on this survey alone would complement other more sensitive and broadband surveys of nearby stars by the {\em Breakthrough Listen} program, and improve upon the current transmitter rate limits at the VLA survey sensitivity by several orders of magnitude. 

We note that the programs sketched out in Figure \ref{fig:seti_rates} represent only a fraction of the operational up-time of the VLA, and commensal observations alongside other primary user observing programs will present ample opportunity for diverse radio technosignature searches across the total frequency range of the VLA.  Moreover, while we have described a program targeting narrow-band emission using conventional incoherent and beamforming techniques, a variety of other powerful search strategies are possible with interferometers \cite{Garrett:2018va}.  We note that the VLA commensal system would be a key asset for addressing many of the science areas highlighted in other Astro2020 Science White Papers, including \cite{berea2020, julia2020, jacob2020}.

The VLA commensal radio SETI system could also be used for a variety of other science cases and adds unique new observational capabilities to the VLA. While the primary purpose of the system is to conduct searches for extraterrestrial intelligence, it can be thought of as essentially an arbitrary high performance digital signal processing system with a full data rate connection to individual antenna voltage data.  In this light, it is clear that it could be applied to a number of compelling use-cases related to fast radio bursts (FRBs). The incredible wealth of scientific opportunities for FRBs are well described in Astro2020 White Papers, including assessing the properties of intergalactic baryons, non-Milky Way intragalactic media, probing dark matter models and exploring fundamental physics in compact objects \cite{law2020, ravi2020, stinebring2020}. Further, the fact that the system could generally act as an additional and complementary observational platform for fast radio transient searches is in and of itself a powerful contribution to FRB science \cite{lynch2020}.  

The current fast transient search and localization system for VLA, {\em realfast} \cite{2018ApJS..236....8L}, acquires data from the current digital infrastructure in a way that limits its temporal and spectral resolution to levels sub-optimal for FRB detection and localization.  Various opportunities exist to improve the performance of matched filter detection for a putative FRB population, including better matching the spectro-temporal properties of the observed bursts \cite{2019arXiv190611476B} and employing machine-learning directly on raw data streams prior to dedispersion \cite{2018ApJ...866..149Z}.  These methods, and others, could readily be implemented with the same data stream and hardware employed with the VLA commensal radio SETI system. In the case of FRBs, an increase in sensitivity, or increase in SNR for constant fluence, brings about not only an increased rate of detection, but also significant improvements in the limits on dedicated follow-up observations and reductions in the localization error ellipse.

\section{Technology Development}
\label{sec:tech}
An Observatory is maximally productive when it has the capability of supporting a diverse range of instruments, each processing data in a way that best matches a particular science goal. Availability of low-level data products -- namely, antenna voltages -- to a dynamic suite of observatory-maintained and user-supplied instruments is key to supporting this diversity.

The mechanism by which data streams are transported in a telescope strongly constrains the extent to which it can be adapted after its initial design. Point-to-point interconnect, such as that based on meshes of cables (as in, for example, ALMA \cite{Escoffier2007}), backplane-technologies (such as those of the Advanced Telecommunications Computing Architecture (ATCA)) , or ``fiber circuitry'' \cite{askap-fiber-interconnect} can make for cost-effective data transport solutions, but, once implemented, define a rigid data routing regime that is difficult to adapt to accommodate new instruments.

In contrast, active-switched data transport mechanisms -- such as Ethernet -- offer the ability to build extensible, flexible, data transport networks with off-the-shelf components. These networks can support one-to-many transmission (multicast) which can enable multiple instruments to observe simultaneously. They may also be trivially extended to support new instruments after a telescope's original inception.

We propose two technology development paths to augment the capabilities of the VLA, and contribute to the design of future facilities, such as the Next-Generation VLA (ngVLA).

\subsection{Adding Ethernet Data Transport to the VLA}
\label{sec:tech:eth}
The first aspect of this proposal is to develop a system capable of copying the voltages generated by the VLA digital front-ends (``station boards'') onto a new 100Gb/s Ethernet switching fabric. This is achieved via the addition of a translator board into the VLA system -- the SETI Wideband Correlator Interface Board (SWIB).

\subsubsection*{\small The SETI Wideband Interface Board}

The SWIB is a non-intrusive interface card which plugs into an unused connector on the VLA ``WIDAR'' correlator's baseline boards \cite{2011-vla, widar}.
This connector provides, in a custom digital format, antenna voltage data for the primary user's selected VLA tuning. This can comprise up to four bands each of 2\,GHz width.

\begin{figure}
    \centering
    \begin{subfigure}[b]{0.8\textwidth}
    \includegraphics[width=\textwidth]{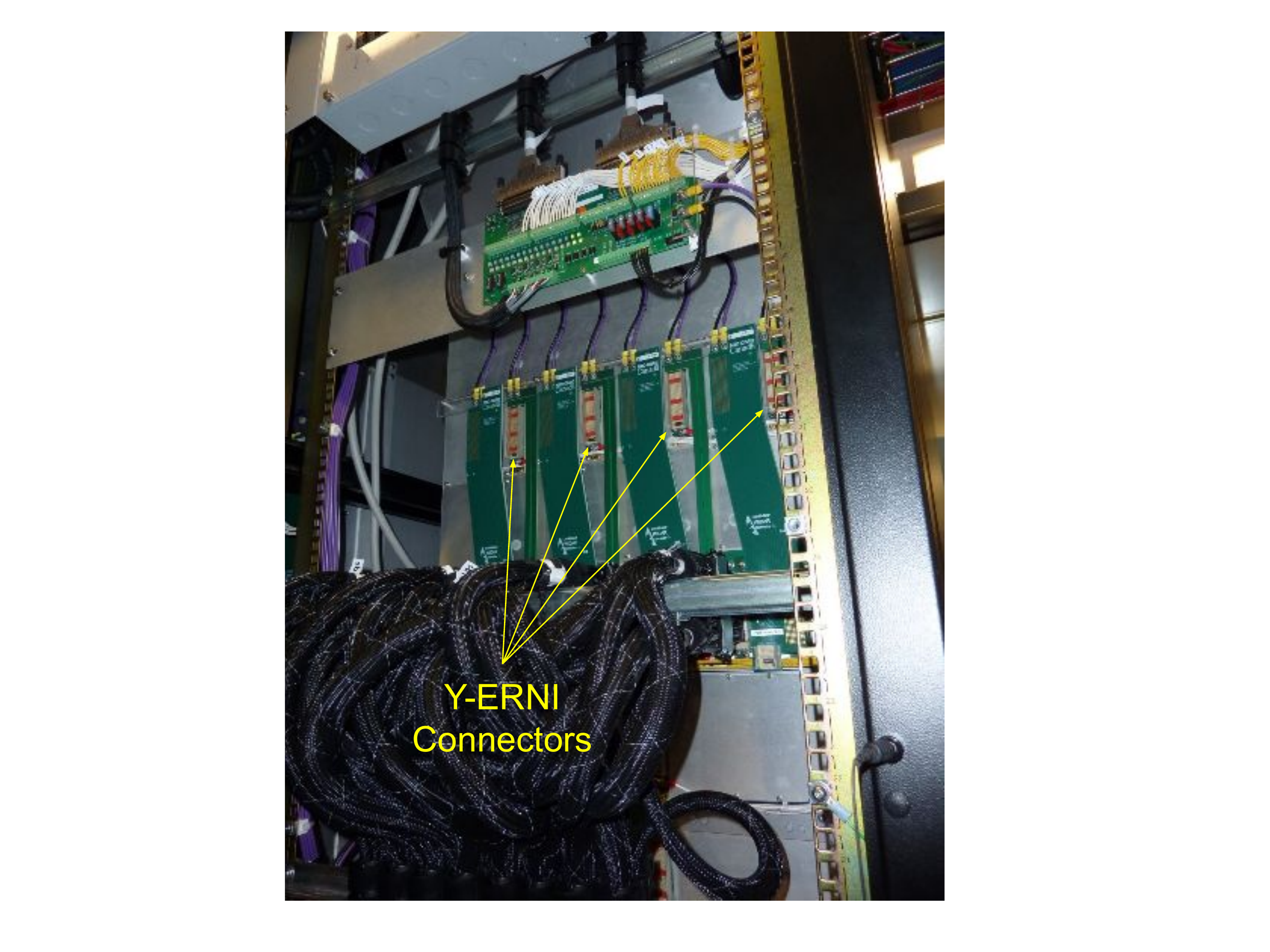}

    \caption{The VLA correlator comprises 64 pairs of baseline boards. Each pair has a spare ``Y-ERNI'' output connector, which can support a SWIB Ethernet interface board}
    \label{fig:y-erni}
    \end{subfigure}
    
    \vspace{0.5cm}
    
    \begin{subfigure}[b]{0.8\textwidth}
    \includegraphics[width=\textwidth]{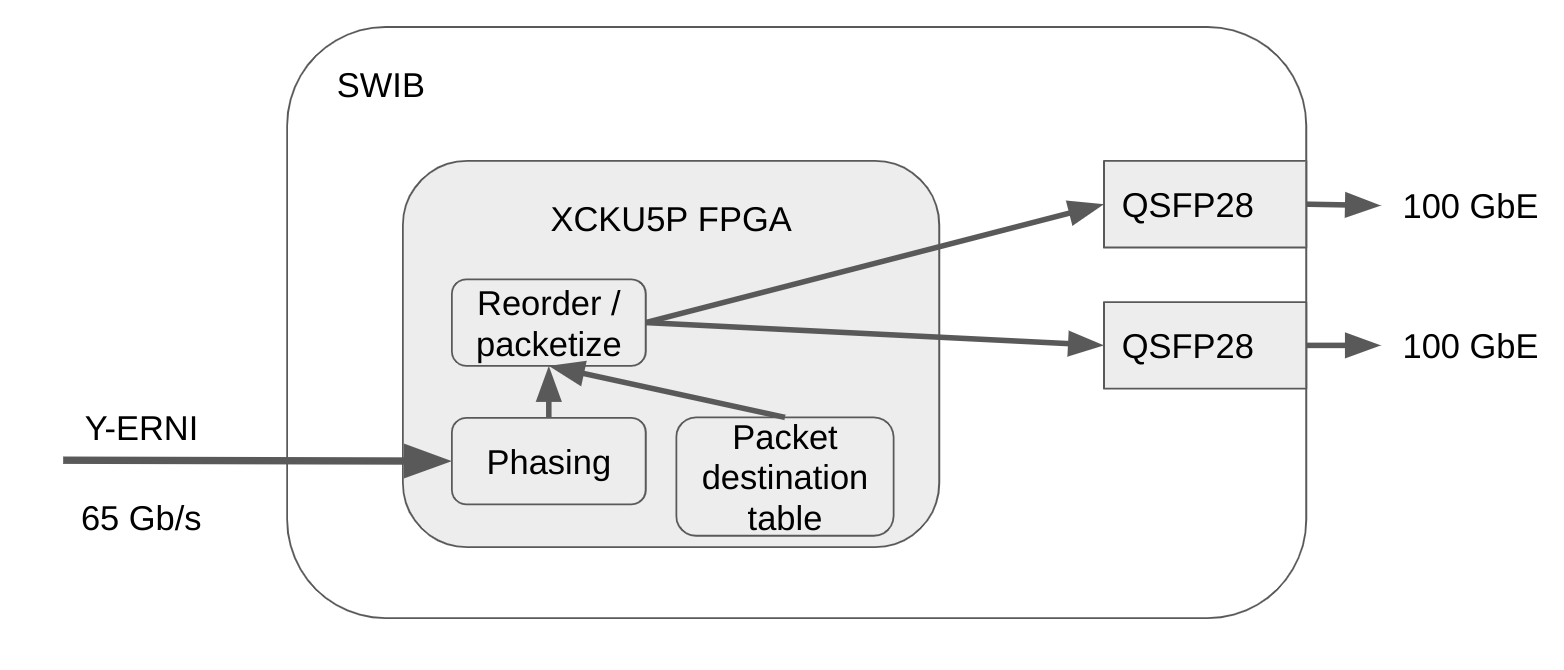}

    \caption{A simple block diagram of the SWIB card, which comprises a ``Y-ERNI'' input, an FPGA which provides phase rotation processing and UDP/IP packetization of data streams onto a pair of 100 Gb/s Ethernet outputs}
    \label{fig:swib}
    \end{subfigure}
    \caption{A SWIB interface card (b) will be added to the spare output connector of each of the VLA correlator's baseline board pairs (a).}
    \label{fig:swib-interface}
\end{figure}

The SWIB -- shown as a functional block diagram in Figure~\ref{fig:swib-interface}, alongside the connectors to which it mates -- is a simple board based around a Field-Programmable Gate Array (FPGA) processor. It is designed to accept antenna voltages as inputs from the VLA baseline boards, perform basic synchronization and phase-rotation operations, and package these data as UDP/IP packet streams. These streams are transmitted over a pair of 100\,Gb Ethernet links to a multi-Tb/s switching network.

The input data rate to each SWIB can be up to 65\,Gb/s, depending on the configuration of the VLA defined by the telescope's primary user. However, the SWIB is provisioned with a pair of 100\,Gb/s Ethernet output ports in order to facilitate bit growth in the planned phasing operation and to allow a user-friendly 8-bit-per-word output format. With these assumptions, the output data rate is $\sim$130\,Gb/s per SWIB.

The complete VLA system comprises 64 pairs of baseline boards. With one SWIB per pair, the proposed system would generate 128 100\,Gb/s Ethernet streams and $\sim$8\,Tb/s of data.

Funding is currently available for design and manufacture of the SWIB board via a private donation to the SETI Institute, with the intention that this board be developed by the VLA correlator's original designers at the National Research Council Canada.

\subsubsection*{\small Ethernet Networking}
Alongside the 128 100\,Gb/s Ethernet sources feeding the proposed network are 128 outputs required to sink the data into processing nodes. These processors are expected to be standard x86-64 GNU/Linux servers, augmented with graphical processing units (GPUs) which are widely used in radio astronomy to accelerate key operations such as channelization \cite{breakthrough-gbt}, correlation \cite{xgpu, chime-xeng}, and beamforming \cite{2015-magro-beamforming}.

100 Gb/s Ethernet technology is already very mature, and currently switches supporting the standard retail at a cost of approximately \$375 per port. This is likely to fall significantly as 200\,Gb and 400\,Gb Ethernet standards become more mainstream, but is already far from the cost driver of the project under consideration.

\begin{figure}
    \centering
    \includegraphics[width=0.5\textwidth]{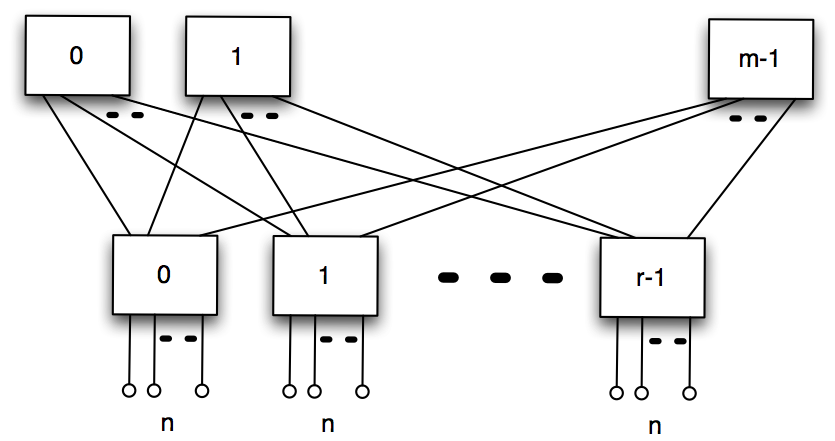}
    \caption{A Clos network is a well-known architecture that allows large switching networks to be built from numerous smaller switching units.}
    \label{fig:clos}
\end{figure}

A switch of the required size (at least 256 ports) to support the full VLA bandwidth is likely to be most effectively implemented using a Clos network (Figure~\ref{fig:clos}) which allows a large network to be constructed from meshes of smaller switching units.
Such an implementation has the disadvantage of requiring more cabling and ports than a single, large, atomic switch. However, the architecture is both flexible and robust and can be easily scaled to meet the demands of a large system.

A Clos architecture has been used in the MeerKAT telescope \cite{2016meerkat, 2018meerkat-nature, manley2015scalable} where, alongside Ethernet's multicast protocol, it enables the telescope to simultaneously service multiple distinct instruments and users. MeerKAT is the first telescope -- but almost certainly will not be the last -- to embrace this paradigm.

\begin{figure}[t]
    \centering
    \includegraphics[width=\textwidth]{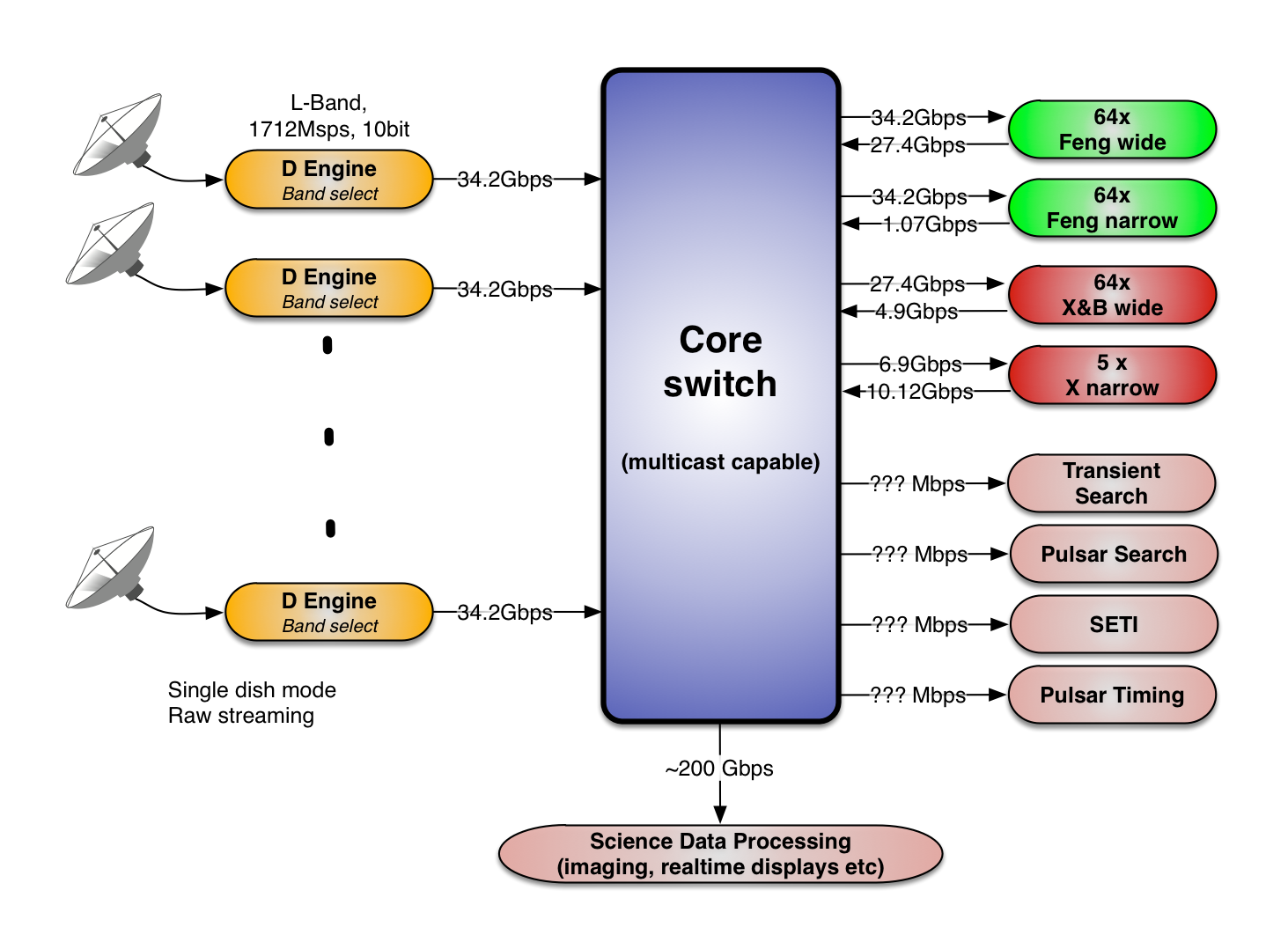}
    \caption{The 40 Gb Ethernet infrastructure used in the MeerKAT array feeds data between 64 digitizers (``D-Engines'') and a variety of downstream processors, including correlators (``X-Engines'') and channelizers (``F-Engines''). The system also allows a variety of users to simultaneously receive data-products generated by the telescope for real-time processing. While the MeerKAT antennas feed the network through 64 network ports, the system provisions 224 network ports ($\sim$9\,Tb/s of bandwidth) for user-supplied equipment.}
    \label{fig:meerkat_40g_connections}
\end{figure}

MeerKAT uses a 40\,Gb Ethernet fabric to interconnect a variety of processing elements -- including channelizers, beamformers, correlators, transient search pipelines, and SETI search pipelines. This interconnect fabric allows users to dynamically subscribe to their choice of data products -- e.g. raw antenna voltages, channelised voltages, visibility matrices, phased-array beams -- depending on the requirements of their science. A schematic depiction of the system is shown in Figure~\ref{fig:meerkat_40g_connections}.

The 64 MeerKAT telescopes generate 2.2 Tb/s of data. However, the system provisions approximately 9 Tb/s of capacity to pass data to user supplied equipment. This is in addition to the interconnection bandwidth used by the observatory's facility correlator/beamformer processors.

That the majority of processing hardware servicing MeerKAT comes from user-supplied equipment serves to illustrate the power of a generic Ethernet-connected architecture. Not only has this architecture allowed users a unique ability to mix their use of facility generated data products with their own processing outputs - the Ethernet architecture has also allowed new users to add unplanned capabilities to the telescope relatively late in its development \cite{2017AcAau.139...98W}.

The multicast-enabled Ethernet-based transport adopted by MeerKAT has long been championed by the NSF-funded Collaboration for Astronomy Signal Processing and Electronics Research (CASPER) \cite{2016JAI.....541001H}.
Adding such a system to the VLA will further cultivate expertise in this arena, to the benefit of future large-scale telescope projects.

The benefits of Ethernet-based interconnect are numerous. Explicitly, some of them are:
\begin{enumerate}
  \item Flexibility: Supports arbitrary numbers of inputs and outputs.
  \item Upgradability: Computing modules and switches themselves can easily be replaced piecemeal to attain lower power consumption or higher performance with relatively low engineering investment. 
  \item Robustness: Spare network ports can be used to easily provision hot-spares and to automatically route data around failed hardware.
  \item Development-friendliness: Features such as multicast, port-mirroring, and industry standard traffic inspection tools allow developers numerous easy ways to verify systems.
  \item Operation-friendliness: Off-the-shelf logging and monitoring tools allow simple health monitoring and statistics gathering for failure analysis and debugging.
  \item Industry Standard: Interfaces to Ethernet networks are supported by all major processing platforms.
  \item Multi-user Support: Natively supports multi-user deployments using multicast subscription.
\end{enumerate}

\subsection{A Commensal SETI System for the VLA}
\label{sec:setivla}

The second aspect of this proposal is to develop and deploy a flexible, high-performance digital signal processing system, to interface with the VLA through the Ethernet interface described in Section~\ref{sec:tech:eth}. This system would be designed as a SETI-focused instrument and would significantly augment the capabilities of the telescope. It would also serve as a platform from which to develop and test software and algorithms for future telescopes.

The proposed system would have full access to the antenna voltage streams from the VLA antennas, and thus -- subject to the bandwidths and pointings chosen by the VLA's primary observer -- its capabilities are effectively limited only by the amount of computing resources dedicated to signal processing.

With computing hardware rapidly evolving over time, the exact specifications of the proposed processing system will be a result of performance benchmarking on hardware purchased as close to deployment as is practicable. However, given the compute-to-IO ratio of modern graphical processing units (GPU) -- in excess of 1000 floating-point operations per byte transferred -- versus that of most basic real-time SETI processing operations associated with a 27-dish array -- the formation of 50 beams requires $\sim$100 floating-point operations per byte transferred -- it is reasonable to assume that the size of the deployed processing system will be dictated by the need to sink up to 8\,Tb/s of data, rather than processing requirements.

A straw-person processing system might consist of 128 dual-CPU-socket x86/64 servers, each hosting between two and four consumer-grade GPU accelerators, and sinking $\sim$64\,Gb/s of raw antenna voltage data.

Buffering of several minutes of data can be achieved by equipping each processing node with $O$(TBytes) of RAM or parallel non-volatile memory express (NVMe) modules. The latter strategy is likely to be far more cost-effective, though such modules have a limited write endurance and would be expected to need replacing at least once a year.

Such a processing system would be capable of a variety of operational modes. Buffering gives the capacity to trade off operating duty cycle against algorithm execution time - the system can either process data in real-time, or, should a more complex algorithm be preferred, can capture a window of several minutes of data and process off-line.

The following operating modes form the base requirements of the processing system:
\begin{itemize}
    \item Real-time generation of $\sim$50 phased-array beams, tracking SETI targets of interest in the VLA's primary beam.
    \item Real-time cross-correlation of all 27 VLA antennas for calibration and fast ($<1$\,ms) imaging.
    \item 1 Hz resolution spectrometry, with multiple beams.
    \item Full-band incoherent beamforming, with dispersion measure searching.
    \item minute-timescale voltage buffering and triggered dumps.
\end{itemize}

Other processing modes for both SETI and non-SETI use cases will likely be explored during the course of commissioning the above capabilities. It is also conceivable that non-GPU accelerators (such as commercial FPGA platforms) might be used to augment the CPU/GPU processing engines where their superior IO-to-compute ratios match to a particular task. 

\subsection{Software development}

Signal processing at a throughput commensurate with the data generation rate of the VLA front-end is undoubtedly a challenge, and is likely several person-years of effort.
However, one of the prime advantages of moving to an Ethernet-driven architecture is that it permits the use of generic CPU and GPU processors for signal processing. In recent years, many software modules have been developed both within and outside the astronomy community which target these platforms, and could be effectively redeployed at the VLA, building a valuable knowledge base to be leveraged when developing future instruments.

\subsection{Broader Impacts}
A critical part of this project is the ability for it to serve as a platform for future astronomers -- in both SETI and non-SETI fields -- to build upon. Flexibility of world-class astronomy instruments to support user-contributed instrumentation is particularly vital at a time when many smaller observatories which support the training of students and early-career researchers are facing significant funding pressures.

In particular, the flexibility and accessibility of this system will provide fertile ground for training the next generation of SETI scientists in world-class fashion, the necessity for which is well described in \cite{wright2020apc}.  This system would also be an ideal development laboratory for designing and testing the methodology required to actualize the significant SETI science opportunities with ngVLA \cite{Croft:2018vv}.

\section{Organization, Partnerships, and Current Status}
Currently, the initial stages of this project are privately funded through the SETI Institute, with memoranda of understanding either in place or in negotiation between the Breakthrough Prize Foundation, NRAO and DRAO.

The concept of Ethernet-based telescope design has long been championed by the NSF-funded Collaboration for Astronomy Signal Processing and Electronics Research (CASPER) \cite{2016JAI.....541001H}, and this project would likely draw in expertise from this collaboration, with significant contributions of software tools and domain expertise from observatories around the world, in particular the South African Radio Astronomy Observatory (SARAO) which supports the MeerKAT array \cite{2016meerkat, 2018meerkat-nature}.

This work will also draw on the open-source software contributions of the privately-funded Breakthrough Listen project, which in recent years has deployed a number of similar scale general-purpose SETI instruments at major world observatories \cite{bl-gbt, bl-parkes}.

\section{Schedule}

\begin{figure}[t]
    \centering
    \includegraphics[width=0.9\textwidth]{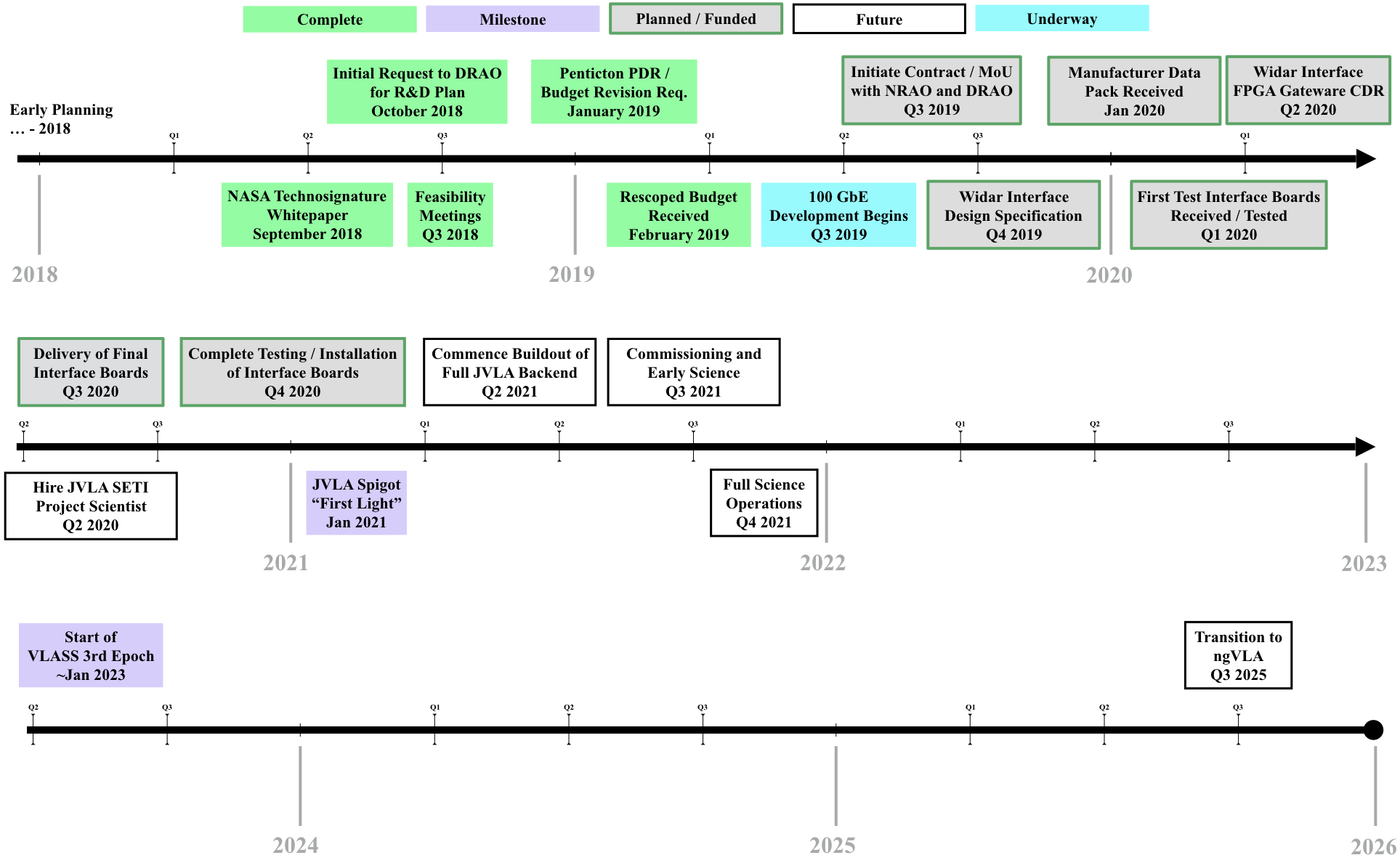}
    \caption{A tentative timeline for the execution of the project and activities described herein, including SWIB design, fabrication, installation, compute cluster installation and leading up to and through the commensal SETI survey described in Figure \ref{fig:seti_rates}.}
    \label{fig:vla_gantt}
\end{figure}

A tentative schedule for this project is shown in Figure~\ref{fig:vla_gantt}. Planning work is currently underway, with fabrication of SWIB modules to take place in 2020. The system is expected to have first light in Q1 2021, with a full-system buildout leading to early science in Q3 2021 and full science operations in Q4 2021. This schedule allows the system to be fully operational for commensal observing with the 3rd epoch of the VLASS survey in 2023.

\section{Cost Estimates}

Full costing breakdown of this pursuit is ongoing. A basic break down of costs, totalling \$9.5M, is shown in Table~\ref{tab:costs}.

These estimates are exclusive of running costs, assumed to be dominated by a power requirement of $~$100\,kW. As is always the case for off-the-shelf computing infrastructure, regular upgrading of hardware can significantly reduce space and power requirements, at the expense of further development costs.

\begin{table}
\begin{center}
\begin{tabular}{c|c}
Item     &  Cost (M\$) \\
\hline \hline
SWIB interface boards (funding available)      &  1.0 \\
Switching Infrastructure            &  0.5 \\
Processing infrastructure           &  4.0 \\
Software development (3 yrs)        &  4.0 \\
\hline
TOTAL                               &  9.5 \\
\hline
\end{tabular}
\end{center}
\caption{A basic breakdown of costs for this small-scale ground-based project.}
\label{tab:costs}
\end{table}

\clearpage
\setcounter{page}{1}
\pagebreak
\bibliographystyle{unsrt}
\bibliography{refs}

\begin{thebibliography}{10}

\bibitem{NAP25252}
Engineering National Academies~of Sciences and Medicine.
\newblock {\em An Astrobiology Strategy for the Search for Life in the
  Universe}.
\newblock The National Academies Press, Washington, DC, 2019.

\bibitem{mercedes2020}
Mercedes Lopez-Morales et~al.
\newblock {Detecting Earth-like Biosignatures on Rocky Exoplanets around Nearby
  Stars with Ground-based Extremely Large Telescopes}.
\newblock {\em Astro2020 Science White Paper}, 2019.

\bibitem{wright2020techno}
Jason Wright.
\newblock {Searches for Technosignatures in Astronomy and Astrophysics}.
\newblock {\em Astro2020 Science White Paper}, 2019.

\bibitem{jlm2020}
Jean-Luc Margot.
\newblock {Radio technosignatures}.
\newblock {\em Astro2020 Science White Paper}, 2019.

\bibitem{2018PNAS..115E9755G}
Claudio Grimaldi and Geoffrey~W Marcy.
\newblock {Bayesian approach to SETI}.
\newblock In {\em Proceedings of the National Academy of Sciences}, pages
  E9755--E9764. Ecole Polytechnique F{\'e}d{\'e}rale de Lausanne
  claudio.grimaldi@epfl.ch; geoff.w.marcy@gmail.com, National Academy of
  Sciences, October 2018.

\bibitem{Lacy:2019uf}
M~Lacy, S~A Baum, C~J Chandler, S~Chatterjee, T~E Clarke, S~Deustua, J~English,
  J~Farnes, B~M Gaensler, N~Gugliucci, G~Hallinan, B~R Kent, A~Kimball, C~J
  Law, T~J~W Lazio, J~Marvil, S~A Mao, D~Medlin, K~Mooley, E~J Murphy, S~Myers,
  R~Osten, G~T Richards, E~Rosolowsky, L~Rudnick, F~Schinzel, G~R Sivakoff,
  L.~O. Sjouwerman, R~Taylor, R~L White, J~Wrobel, A~J Beasley, E~Berger,
  S~Bhatnagar, M~Birkinshaw, G~C Bower, W~N Brandt, S~Brown, S~Burke-Spolaor,
  B~J Butler, J~Comerford, P~B Demorest, H~Fu, S~Giacintucci, K~Golap, T~Guth,
  C~A Hales, R~Hiriart, J~Hodge, A~Horesh, Z~Ivezic, M~J Jarvis, A~Kamble,
  N~Kassim, X~Liu, L~Loinard, D~K Lyons, J~Masters, M~Mezcua, G~A Moellenbrock,
  T~Mroczkowski, K~Nyland, C~P O'Dea, S~P O'Sullivan, W~M Peters, K~Radford,
  U~Rao, J~Robnett, J~Salcido, Y~Shen, A~Sobotka, S~Witz, M~Vaccari, R~J van
  Weeren, A~Vargas, P~K~G Williams, and I~Yoon.
\newblock {The Karl G. Jansky Very Large Array Sky Survey (VLASS). Science
  case, survey design and initial results}.
\newblock {\em arXiv.org}, July 2019.

\bibitem{Garrett:2018va}
M~A Garrett.
\newblock {SETI surveys of the nearby and distant universe employing wide-field
  radio interferometry techniques}.
\newblock {\em arXiv.org}, October 2018.

\bibitem{berea2020}
Anamaria Berea.
\newblock {Applications of data science}.
\newblock {\em Astro2020 Science White Paper}, 2019.

\bibitem{julia2020}
Julia DeMarines et~al.
\newblock {Observing the Earth as a communicating exoplanet}.
\newblock {\em Astro2020 Science White Paper}, 2019.

\bibitem{jacob2020}
Jacob Haqq-Misra et~al.
\newblock {Scientific implications of (non-)detection}.
\newblock {\em Astro2020 Science White Paper}, 2019.

\bibitem{law2020}
Casey Law et~al.
\newblock {Radio Time-Domain Signatures of Magnetar Birth}.
\newblock {\em Astro2020 Science White Paper}, 2019.

\bibitem{ravi2020}
Vikram Ravi et~al.
\newblock {Fast Radio Burst Tomography of the Unseen Universe}.
\newblock {\em Astro2020 Science White Paper}, 2019.

\bibitem{stinebring2020}
Dan Stinebring et~al.
\newblock {Twelve Decades: Probing the Interstellar Medium from kiloparsec to
  sub-AU scales}.
\newblock {\em Astro2020 Science White Paper}, 2019.

\bibitem{lynch2020}
Ryan Lynch et~al.
\newblock {The Virtues of Time and Cadence for Pulsars and Fast Transients}.
\newblock {\em Astro2020 Science White Paper}, 2019.

\bibitem{2018ApJS..236....8L}
C~J Law, G~C Bower, S~Burke-Spolaor, B~J Butler, P~Demorest, A~Halle,
  S~Khudikyan, T~J~W Lazio, M~Pokorny, J~Robnett, and M~P Rupen.
\newblock {realfast: Real-time, Commensal Fast Transient Surveys with the Very
  Large Array}.
\newblock {\em The Astrophysical Journal Supplement Series}, 236(1):8, May
  2018.

\bibitem{2019arXiv190611476B}
K~W Bannister, A~T Deller, C~Phillips, J-P Macquart, J~X Prochaska, N~Tejos,
  S~D Ryder, E~M Sadler, R~M Shannon, S~Simha, C~K Day, M~McQuinn, F~O
  North-Hickey, S~Bhandari, W~R Arcus, V~N Bennert, J~Burchett, M~Bouwhuis,
  R~Dodson, R~D Ekers, W~Farah, C~Flynn, C~W James, M~Kerr, E~Lenc, E~K Mahony,
  J~O'Meara, S~Os{\l}owski, H~Qiu, T~Treu, V~U, T~J Bateman, D~C-J Bock, R~J
  Bolton, A~Brown, J~D Bunton, A~P Chippendale, F~R Cooray, T~Cornwell,
  N~Gupta, D~B Hayman, M~Kesteven, B~S Koribalski, A~MacLeod, N~M
  McClure-Griffiths, S~Neuhold, R.~P. Norris, M~A Pilawa, R~Y Qiao, J~Reynolds,
  D~N Roxby, T~W Shimwell, M~A Voronkov, and C~D Wilson.
\newblock {A single fast radio burst localized to a massive galaxy at
  cosmological distance}.
\newblock {\em arXiv.org}, page arXiv:1906.11476, June 2019.

\bibitem{2018ApJ...866..149Z}
Yunfan~Gerry Zhang, Vishal Gajjar, Griffin Foster, Andrew Siemion, James
  Cordes, Casey Law, and Yu~Wang.
\newblock {Fast Radio Burst 121102 Pulse Detection and Periodicity: A Machine
  Learning Approach}.
\newblock {\em The Astrophysical Journal}, 866(2):149, October 2018.

\bibitem{Price:2019tg}
Danny~C Price, J~Emilio Enriquez, Bryan Brzycki, Steve Croft, Daniel Czech,
  David DeBoer, Julia DeMarines, Griffin Foster, Vishal Gajjar, Nectaria
  Gizani, Greg Hellbourg, Howard Isaacson, Brian Lacki, Matt Lebofsky, David
  H~E MacMahon, Imke de~Pater, Andrew P.~V. Siemion, Dan Werthimer, James~A
  Green, Jane~F Kaczmarek, Ronald~J Maddalena, Stacy Mader, Jamie Drew, and
  S~Pete Worden.
\newblock {The Breakthrough Listen Search for Intelligent Life: Observations of
  1327 Nearby Stars over 1.10-3.45 GHz}.
\newblock {\em arXiv.org}, June 2019.

\bibitem{Escoffier2007}
{Escoffier, R. P.}, {Comoretto, G.}, {Webber, J. C.}, {Baudry, A.}, {Broadwell,
  C. M.}, {Greenberg, J. H.}, {Treacy, R. R.}, {Cais, P.}, {Quertier, B.},
  {Camino, P.}, {Bos, A.}, and {t, A. W. Gun}.
\newblock The {ALMA} correlator.
\newblock {\em A\&A}, 462(2):801--810, 2007.

\bibitem{askap-fiber-interconnect}
G.~Hampson, A.~Brown, S.~Neuhold, J.~Bunton, A.~Macleod, J.~Tuthill, and
  R.~Beresford.
\newblock Askap advancements in beamformer and correlator optical backplane
  technology.
\newblock In {\em 2013 US National Committee of URSI National Radio Science
  Meeting (USNC-URSI NRSM)}, pages 1--1, Jan 2013.

\bibitem{2011-vla}
S.~M. {Dougherty} and Rick {Perley}.
\newblock {The Expanded Very Large Array}.
\newblock {\em Bulletin de la Societe Royale des Sciences de Liege},
  80:491--495, Jan 2011.

\bibitem{widar}
Brent Carlson.
\newblock {Refined EVLA WIDAR Correlator Architecture}.
\newblock {\em NRC EVLA Memo Series}, 4, 10 2001.

\bibitem{breakthrough-gbt}
David H.~E. {MacMahon}, Danny~C. {Price}, Matthew {Lebofsky}, Andrew P.~V.
  {Siemion}, Steve {Croft}, David {DeBoer}, J.~Emilio {Enriquez}, Vishal
  {Gajjar}, Gregory {Hellbourg}, and Howard {Isaacson}.
\newblock {The Breakthrough Listen Search for Intelligent Life: A Wideband Data
  Recorder System for the Robert C. Byrd Green Bank Telescope}.
\newblock {\em Publications of the Astronomical Society of the Pacific},
  130(986):044502, Apr 2018.

\bibitem{xgpu}
M.~A. {Clark}, P.~C. {La Plante}, and L.~J. {Greenhill}.
\newblock {Accelerating Radio Astronomy Cross-Correlation with Graphics
  Processing Units}.
\newblock {\em arXiv e-prints}, page arXiv:1107.4264, Jul 2011.

\bibitem{chime-xeng}
Nolan {Denman}, Mandana {Amiri}, Kevin {Bandura}, Jean-Fran{\c{c}}ois {Cliche},
  Liam {Connor}, Matt {Dobbs}, Mateus {Fandino}, Mark {Halpern}, Adam {Hincks},
  and Gary {Hinshaw}.
\newblock {A GPU-based Correlator X-engine Implemented on the CHIME
  Pathfinder}.
\newblock {\em arXiv e-prints}, page arXiv:1503.06202, Mar 2015.

\bibitem{2015-magro-beamforming}
A.~{Magro}, K.~Zarb {Adami}, and J.~{Hickish}.
\newblock {GPU-Powered Coherent Beamforming}.
\newblock {\em Journal of Astronomical Instrumentation}, 4:1550002--336, Jun
  2015.

\bibitem{2016meerkat}
{\em {MeerKAT Science: On the Pathway to the SKA}}, Jan 2016.

\bibitem{2018meerkat-nature}
Fernando {Camilo}.
\newblock {African star joins the radio astronomy firmament}.
\newblock {\em Nature Astronomy}, 2:594--594, Jul 2018.

\bibitem{manley2015scalable}
Jason~Ryan Manley.
\newblock {\em A scalable packetised radio astronomy imager}.
\newblock PhD thesis, University of Cape Town, 2015.

\bibitem{2017AcAau.139...98W}
S~Pete Worden, Jamie Drew, Andrew Siemion, Dan Werthimer, David DeBoer, Steve
  Croft, David MacMahon, Matt Lebofsky, Howard Isaacson, Jack Hickish, Danny
  Price, Vishal Gajjar, and Jason~T. Wright.
\newblock {Breakthrough Listen - A new search for life in the universe}.
\newblock {\em Acta Astronautica}, 139:98--101, October 2017.

\bibitem{2016JAI.....541001H}
Jack {Hickish}, Zuhra {Abdurashidova}, Zaki {Ali}, Kaushal~D. {Buch},
  Sandeep~C. {Chaudhari}, Hong {Chen}, Matthew {Dexter}, Rachel~Simone
  {Domagalski}, John {Ford}, and Griffin {Foster}.
\newblock {A Decade of Developing Radio-Astronomy Instrumentation using CASPER
  Open-Source Technology}.
\newblock {\em Journal of Astronomical Instrumentation}, 5(4):1641001--12, Dec
  2016.

\bibitem{wright2020apc}
Jason Wright.
\newblock {SETI State of the Profession}.
\newblock {\em Astro2020 APC White Paper}, 2019.

\bibitem{Croft:2018vv}
Steve Croft, Andrew P.~V. Siemion, James~M Cordes, Ian~S Morrison, Zsolt
  Paragi, Jill Tarter, , U~C Berkeley, {Radboud University}, {SETI Institute},
  {Cornell University}, {Curtin University}, and {JIVE}.
\newblock {Science with an ngVLA: SETI Searches for Evidence of Intelligent
  Life in the Galaxy}.
\newblock {\em arXiv.org}, October 2018.

\bibitem{bl-gbt}
David H.~E. {MacMahon}, Danny~C. {Price}, Matthew {Lebofsky}, Andrew P.~V.
  {Siemion}, Steve {Croft}, David {DeBoer}, J.~Emilio {Enriquez}, Vishal
  {Gajjar}, Gregory {Hellbourg}, and Howard {Isaacson}.
\newblock {The Breakthrough Listen Search for Intelligent Life: A Wideband Data
  Recorder System for the Robert C. Byrd Green Bank Telescope}.
\newblock {\em Publications of the Astronomical Society of the Pacific},
  130(986):044502, Apr 2018.

\bibitem{bl-parkes}
Danny~C. Price, David H.~E. MacMahon, Matt Lebofsky, Steve Croft, David DeBoer,
  J.~Emilio Enriquez, Griffin~S. Foster, Vishal Gajjar, Nectaria Gizani, Greg
  Hellbourg, and et~al.
\newblock The breakthrough listen search for intelligent life: Wide-bandwidth
  digital instrumentation for the csiro parkes 64-m telescope.
\newblock {\em Publications of the Astronomical Society of Australia}, 35:e041,
  2018.

\end{thebibliography}

\end{document}